\documentstyle [12pt] {article}
\def\pr#1#2#3{{\em Phys. Rev.} {\bf D#1} (19#3) #2}

\def\a{\alpha }

\begin{document}
\begin{center}
{\Large{\bf{Measurement of $xF_3$, $F_2$ Structure Functions \\
and Gross--Llewellyn Smith Sum Rule \\
with IHEP--JINR  Neutrino Detector\\}}}
\end{center}

\vspace{0.5cm}
\begin{center}
L.~S.~Barabash, ~S.~A.~Baranov, ~Y.~A.~Batusov, ~S.~A.~Bunyatov,\\
O.~Y.~Denisov, ~M.~Y.~Kazarinov, ~O.~L.~Klimov, ~A.~V.~Krasnoperov,\\
V.~E.~Kuznetsov, ~V.~V.~Lyukov, ~Y.~A.~Nefedov, ~B.~A.~Popov, ~S.~N.~Prakhov,\\
A.~V.~Sidorov, ~V.~I.~Snyatkov, ~V. V. Tereshchenko, ~V.~Y.~Valuev\\
\vspace{0.2cm}
{\footnotesize{Joint Institute for Nuclear Research, Dubna, Russia\footnote
{Supported by
the Russian Foundation for Basic Research, Grant No 96-02-18562,
by INTAS under Contract No 93-1180 and by
the Russian Foundation for Basic Research, Grant No 96-02-18897.}}
}

\vspace{0.7cm}
V.~B.~Anykeyev, ~A.~A.~Borisov, ~N.~I.~Bozhko, ~S.~K.~Chernishenko,\\
G.~L.~Chukin, ~R.~M.~Fachrutdinov, ~V.~N.~Goryachev,\\
M.~M.~Kirsanov, ~A.~I.~Kononov, ~A.~S.~Kozhin, ~V.~I.~Kravtsov,\\
V.~V.~Lipajev, ~A.~I.~Mukhin, ~S.~A.~Mukhin, ~Y.~I.~Salomatin, \\
Y.~M.~Sapunov, ~K.~E.~Shestermanov, ~A.~A.~Spiridonov, \\
~Y.~M.~Sviridov, ~V.~V.~Sytnik, ~V.~L.~Tumakov, ~A.~S.~Vovenko\\
\vspace{0.2cm}
{\footnotesize{Institute for High Energy Physics, Protvino, Russia\footnote
{Supported by
the Russian Foundation for Basic Research, Grant No 96-02-17608.
}
\\}}
\end{center}
\vspace{0.5cm}

\begin{abstract}
The isoscalar structure functions ~$xF_3$~ and ~$F_2$~ are measured
as functions of ~$x$~ averaged over all ~$Q^2$ permissible
for the range ~$6$~ to ~$28\ GeV$ of incident (anti)neutrino energy.
With the measured values of ~$xF_3$,
the value of the Gross--Llewellyn Smith sum rule is found to be
~$\int_{0}^{1}{F_3\,dx}=2.13\pm0.38\,(stat)\pm0.26\,(syst)$.
The QCD analysis of ~$xF_3$~ provides
$\Lambda_{\overline{MS}}$=358$\pm$59 $MeV$.
The obtained value of the strong interaction constant
~$\alpha_S (M_Z)=0.120^{+3}_{-4}$
is larger than most of the deep inelastic scattering results.
\end{abstract}

%
\newpage
\par
\vspace{1cm}
      The data on deep-inelastic $\nu(\overline{\nu})$--scattering
in a wide region of momentum transfer provide a reliable basis
for precise verification of QCD predictions~\cite{Altarelli}.
In this paper the data on the ~$xF_3$~ and ~$F_2$~
structure functions ~(SF)~ are presented for the
kinematic region of relatively small momentum transfer
~$0.55<Q^2<4.0\,GeV^2$. ~The value of the Gross--Llewellyn Smith
(GLS) sum rule~\cite{gls} and ~$\alpha_S\,(M_Z)$~ are evaluated.
\par
      The data samples were obtained from three independent exposures
of the IHEP--JINR Neutrino Detector~\cite{ND} to the wide band neutrino
and antineutrino beams~\cite{beams} of the Serpukhov U70 accelerator.
The exposure to the antineutrino beam ($\bar \nu_\mu$-exposure)
was performed at the proton beam energy
$E_p=70\,GeV$, whereas the two $\nu_\mu$-exposures
were carried out one at $E_p=70\,GeV$ and the other at $E_p=67\,GeV$.
The experimental set--up and selection criteria for CC
events are discussed in~\cite{publ}.
We restricted the range of the measurements in
$W^2$ to $W^2 > 1.7 GeV^2$ and in $E_{\nu(\overline\nu)}$~
to ~$6 < E_{\nu(\overline\nu)} < 28 GeV$.
~The final number of events and the mean values of ~$Q^2$,
~$\langle Q^2 \rangle$ ~for the three samples are given in Table~1.
\par
    The SF were measured as functions of $x$
averaged over all $Q^2$ permissible for the energy
range ~$6<E_{\,\nu\,(\overline\nu)}<28\,GeV$.
Events were binned in intervals
of $x$, and values of $xF_3$ and $F_2$ were calculated in
these intervals.
\par
    The number of $\nu_\mu$ interactions,
$n^\nu$, and $\overline\nu_\mu$ interactions, $n^{\overline\nu}$, in a
given bin of $x$ is a linear combination of the 'average' values
~$\{F_2\}$~ and ~$\{xF_3\}$~ of the respective SF in this bin
(we assume invariance under the charge conjugation):

$$n^{\,\overline\nu} ~=~ a^{\,\overline\nu}\cdot \{F_2\}
   ~-~ b^{\,\overline\nu}\cdot\{xF_3\}$$

$$n^{\,\nu}_{\,1,2} ~=~
        a^{\,\nu}_{\,1,2}\cdot \{F_2\} ~+~ b^{\,\nu}_{\,1,2}
        \cdot \{xF_3\}.$$

The subscripts $1$ and $2$ correspond to the $\nu_\mu$-exposures at
$E_p=70\,GeV$ and $E_p=67\,GeV$ respectively. The quantities
~$a^{\,\nu,\,\overline\nu}$~ and ~$b^{\,\nu,\,\overline\nu}$~
are integrals ('flux integrals')
of products of the differential neutrino (antineutrino)
flux $\phi^{\,\nu\,(\overline\nu)}\,(E)$ and known factors
depending on the scaling variables ~$x$, $y$~ as foreseen by the
standard form of the differential cross-section for deep-inelastic
$\nu_\mu\,(\overline\nu_\mu)$-scattering off an isoscalar target:

$$a^{\,\overline\nu}~=~
    N\, \frac{G^2M}{\pi}
    \int_{}^{}(1-y-\frac{Mxy}{2E}+\frac{1}{2\,(R+1)}\,y^2)\,
    E\,\phi^{\,\overline\nu}(E)\,dx\,dy\,dE\,, $$

$$b^{\,\overline\nu}~=~
    N\, \frac{G^2M}{\pi}
    \int_{}^{}y\,(1-\frac{y}{2})\,E\,\phi^{\,\overline\nu}
    (E)\,dx\,dy\,dE $$
etc.
Here $N$ is the number of nucleons in the fiducial volume
of the detector and the parameter ~$R=(F_2-2xF_1)\Big/2xF_1$~ measures
the violation of Callan-Gross relation~\cite{callan}.
\par
The number $n_{\,\nu(\overline\nu)}$ of
neutrino (antineutrino) interactions in a given $x$-bin was
obtained from the measured number of neutrino (antineutrino) events
in this bin corrected for acceptance, for
smearing effects arising from Fermi motion and
measurement uncertainties, for radiative effects (following the
prescription given by De R\'ujula et al.~\cite{rujula}) and for
target non-isoscalarity (assuming $d_v\Big/u_v=0.5$~\cite{noni}).
To determine the appropriate correcting factors the Monte--Carlo
simulation of the experimental set--up has been carried out using
the program CATAS~\cite{catas}.
We used the Buras and Gaemers (BEBC)
parametrization~\cite{buras} for quark distributions. The charmed
quark content of the nucleon was assumed to be zero. The kinematic
suppression of $d\rightarrow c$ and $s\rightarrow c$ transitions
was taken into account assuming slow rescaling~\cite{slow} and
the following charmed and strange quark masses:
$m_c=1.25\,GeV$, $m_s=0.25\,GeV$. Fermi motion of
nucleons was simulated according to~\cite{fermi}.
The details of the Monte--Carlo
simulation of the known features of the experimental set--up are
discussed in~\cite{publ} and~\cite{blum}.
\par
The number of interactions in a given bin of $x$ is subject to
kinematic constraints imposed by the cuts in the muon
momentum ($P_\mu > 1\,GeV/c$~~\cite{publ}),
in the neutrino energy ($6<E_{\,\nu,\overline\nu}<28\,GeV$) and in the
invariant mass square of the hadronic system
($W^2 > 1.7\,GeV^2$).
These were taken into account in the
calculation of the flux integrals by appropriate modification of the
volume of integration.
\par
The measured values of ~$xF_3$~ and ~$F_2$~ are presented in
Table~2 and in Figure~1.

The systematic errors presented come from the uncertainties
of the correcting factors due to the choice of some input
quark distributions in the event simulation program CATAS.
These systematic uncertainties were estimated by repeating
the calculation of the SF using
by turns the Field-Feynman~\cite{field} and GRV~\cite{grv}
quark distributions.
Note that the systematic errors in Table~2
do not include the normalization error of $4\%$ for
$F_2$ and $11\%$ for $xF_3$. These normalization errors
originate from the uncertainties in the $\nu_\mu$
and $\overline\nu_\mu$ flux determination~\cite{flux}.


\par
    With the values of $xF_3$, the GLS sum rule ( the integral of $F_3$ )
has been estimated.
Over the interval $0.02 < x < 0.65$ it was calculated by numerical
integration of the measured values of $xF_3$ weighted by $1\Big/ x$.
The contribution from the regions $0 < x < 0.02$ and $0.65 < x < 1$
was evaluated by integrating over these regions the parametrization
of $xF_3$ with the values of free parameters obtained from the fit
to the data at $0.02 < x < 0.65$.
Finally we obtained
\begin{equation}
\int_{0}^{1}\frac{xF_3(x)}{x}\,dx=
    2.13\pm0.38\,(stat)\pm0.26\,(syst).
\label{glsser}
\end{equation}
The systematic error quoted is the quadrature sum of $\pm0.24$
due to $\nu_\mu$ and $\overline\nu_\mu$ flux uncertainties, and $\pm0.09$
due to the choice of some input quark distributions.
In accordance with Table~1 we suppose that
the measured value~(\ref{glsser}) of the GLS sum rule corresponds to
the averaged value ~$\overline{Q^2}\sim~1.7\,GeV^2$.
\par
    The experimental data on the $xF_3$ were compared with
the QCD prediction for $Q^2$-evolution
by the Jacobi polynomials method in the
next-to-leading order QCD approximation \cite{Jacobi,Kriv,Kriv2,KaSi}~.
Making QCD analysis of the $xF_3$ ~SF, for the first step
we do not discuss the problem of validity
of application of perturbative QCD predictions for kinematical
region of small $Q^2$ as well as the nuclear effects, heavy quarks
threshold effects and higher order QCD corrections.
\par
    In order to take into account the target mass corrections
the Nachtmann moments~\cite{Nacht} of $F_3$ and $F_2$
could be expanded in powers of $M_{nucl.}^2/Q^2$,
and retaining only  terms of the order $M_{nucl.}^2/Q^2$
one could obtain:
\begin{eqnarray}
M_{3(2)}(N,Q^2)=M_{3(2)}^{QCD}(N,Q^2)\,+\,
\frac{N(N{+\atop {(-)}}1)}{N+2}\,\frac{M_{nucl.}^2}{Q^2}
\,M_{3(2)}^{QCD}(N+2,Q^2).
\label{m3}
\end{eqnarray}
Here $M_3^{QCD}(N,Q^2)$ and $ M_2^{QCD}(N,Q^2)$ are the Mellin moments of
$xF_3$ and $F_2$:
\begin{eqnarray}
M_3(N,Q^2)&=&\int_{0}^{1}dx{x^{N-2}}xF_{3}(x,Q^2),\nonumber \\
{}\label{Mellf}\\
M_2(N,Q^2)&=&\int_{0}^{1}dx{x^{N-2}}F_{2}(x,Q^2), ~~~~~N = 2,3, ...
                                                  \nonumber
\end{eqnarray}
\par
  The $Q^2$-evolution of $ M_3^{QCD}(N,Q^2)$ and $M_2^{QCD}(N,Q^2)$
is defined ~\cite{s4,s5} by QCD and is presented here for the
nonsinglet case for simplicity:
\begin{eqnarray}
M_3^{QCD}(N,Q^2)
& =& \left [ \frac{\alpha _{S}\left ( Q_{0}^{2}\right )}
{\alpha _{S}\left ( Q^{2}\right )}\right ]^{d_{N}}
H_{N}\left (  Q_{0}^{2},Q^{2}\right )
M_3^{QCD}(N,Q^2_0) ,~~~N = 2,3, ... \label{m3q2} \\
{}\nonumber \\
d_N & = & \gamma^{(0),N}\Bigg/2\beta_0,
        \nonumber
\end{eqnarray}
 Here ~$\a_s(Q^2)$~ is the          strong interaction constant,
~$\gamma^{(0)NS}_{N}$~ are the nonsinglet leading order anomalous dimensions,
and the factor ~$H_{N}\left (  Q_{0}^{2},Q^{2}\right )$~ contains all
next-to-leading order QCD corrections~\cite{KaSi,s5,KKPS}.
\par
The unknown coefficients $M_3(N,Q^2_0)$ in (\ref{m3q2}) could be parametrized
as the Mellin moments of some function:
\begin{eqnarray}
M_3^{QCD}(N,Q^2_0)&=&\int_{0}^{1}dx{x^{N-2}}Ax^b(1-x)^c(1+\gamma x),
~~~ N = 2,3, ...
\label{Mellf30}
\end{eqnarray}
where the constants ~A, b, c and $\gamma$ should be determined
>from the fit to the data.
Having at hand the moments (\ref{m3})$\,$--$\,$(\ref{Mellf30})
and following the method discussed in \cite{Jacobi,Kriv,Kriv2}
we can write the $xF_3$ ~SF in the form:
$$
xF_{3}^{N_{max}}(x,Q^2)=x^{\a}(1-x)^{\beta}\sum_{n=0}^{N_{max}}
\Theta_n ^{\a , \beta}
(x)\sum_{j=0}^{n}c_{j}^{(n)}{(\a ,\beta )}
M_3^{QCD} \left ( j+2,\,Q^{2}\right ),
$$
where $~\Theta^{\alpha \beta}_{n}(x)~$ are the Jacobi polynomials
and $~c^{n}_{j}(\alpha,\beta)~$ are the coefficients of the
expansion of $~\Theta^{\alpha,\beta}_{n}(x)~$ in powers of x:
$$
\Theta_{n} ^{\a , \beta}(x)=
\sum_{j=0}^{n}c_{j}^{(n)}{(\a ,\beta )}\,x^j\,.
$$
The accuracy of the SF approximation better than $~10^{-3}~$ is achieved for
$~N_{max} = 12~$ in a wide region of the
parameters $~ \alpha~$ and $~\beta~$~ \cite{Kriv,Kriv2}.
\par
    Using  nine Mellin moments for SF reconstruction
and taking into account target mass corrections we have
determined five free parameters
A, b, c, $\gamma$ and the QCD parameter $\Lambda_{\overline{MS}}$
(Table~3).

\par
For the $Q^2$~--~dependence of the GLS sum rule we can write the following
theoretical expression\footnote{See \cite{KatSt} for higher order QCD
corrections to the GLS sum rule.} :
\begin{equation}
GLS(Q^2)  =  3\left[1-\alpha_s(Q^2)/\pi + O(\alpha_s^2)
-\frac{8}{27}\frac{\langle\langle O\rangle\rangle}{Q^2}\right]
         \label{resqcd}
\end{equation}
where $\alpha_s$ is the coupling constant in the
$\overline{MS}$ scheme.
The general structure of the high-twist (HT) term is known
from~\cite{SV}.
The evaluation of this term was carried out in \cite{BK},
~$\langle\langle O\rangle\rangle=0.33\pm0.16\ GeV^2$,
and more recently in \cite{HT},
~$\langle\langle O\rangle\rangle=0.53\pm0.04\ GeV^2$,
using the same three-point function QCD sum
rules technique. In order to estimate the uncertainties due to the HT
contribution we included a fenomenological term\footnote{
This shape of h(x) is in qualitative agreement with the theoretical prediction
in \cite{webber} and experimental estimations in \cite{Siht} for the $x$
values from Table~2.}
~$h(x)=-\frac{8}{27}\frac{\langle\langle O\rangle\rangle}{Q^2}~x$
in the fitting procedure .
The first moment of the function $h(x)$ gives some contribution
to the GLS sum rule (\ref{resqcd}) in accordance with \cite{HT}.
The results of the fit with
~$\langle\langle O\rangle\rangle=0.53\pm0.04\ GeV^2$
are presented in Table~3.
The value of $\alpha_S (M_Z)$ was calculated for both variants of the fit
due to the so--called `matching relation'~~\cite{marc}.
We present the GLS sum rule values calculated through (\ref{Mellf30})
with $N=1$ and with the parameters from Table~3.
\par
We repeated our fit taking into account both the statistical and
systematic errors (from Table~2 added in quadrature.
With HT from \cite{HT}, the following estimations have been obtained:
$\Lambda_{\overline{MS}}=359 \pm 71~MeV$, ~$GLS=2.66$~.
\par
In the singlet case the moments of valence quarks, sea quarks and gluons
were parametrized at $Q^2_0$ in the form:
\begin{eqnarray}
M_q^{QCD}(N,Q^2_0)&=&\int_{0}^{1}dx{x^{N-2}}[A_vx^{b_v}(1-x)^{c_v}
+A_{sea}(1-x)^{c_{sea}}],    \nonumber \\
{}\label{Mellfff} \\
M_g^{QCD}(N,Q^2_0)&=&\int_{0}^{1}dx{x^{N-2}}A_g(1-x)^{c_g},
~~~~~N = 2,3, ... \nonumber
\end{eqnarray}
Keeping in mind the small number of experimental points we fix
$A_g$ from the momentum sum rule $M_q^{QCD}(2,Q^2)+ M_g^{QCD}(2,Q^2)=1$~.
Following the results \cite{Kriv2} of the QCD analysis of $F_2$ at
the momentum transfer  $Q^2=5~GeV^2$
we put $A_{sea}=0.17$,~$c_{sea}=15$ and $c_g=9$.
The other parameters in (\ref{Mellfff}) as well as $\Lambda$
were determined from the fit of the data in the leading logarithm QCD
approximation and were found to be
~$  A_v=2.49 \pm 0.311$,
~$ b_v=0.19  \pm 0.02$,
~$ c_v=2.80  \pm 0.05$,
~$ \Lambda=(517  \pm 17)\,MeV$ with $\chi^2=6.7$ for 6 experimental points
and $Q^2_0=3~GeV^2$. Only statistical errors were taken into account.

\newpage
     Several comments:
\begin{itemize}
\item The values (\ref{glsser}) and the results on the GLS sum rule
      in Table~3 are considerably smaller
      in comparison to the results of previous measurements.
      (See the summary on the GLS sum rule data in~\cite{tvse}
      and the latest 3-loop result~\cite{alzccfr}.)
\item The parameter $\Lambda_{\overline{MS}}$ is found to be about twice
      as large as the estimations  in \cite{KaSi} and  \cite{Quintas}.
      It is in qualitative agreement with the results of the NLO
      analysis~\cite{ChK} of the GLS sum rule in the $\overline{MS}$
      scheme:
   ~$\Lambda_{\overline{MS}}^{(4)}=317\pm23(stat)\pm99(syst)\pm62(twist)\,MeV$
      with HT and
   ~$\Lambda_{\overline{MS}}^{(4)}=435\pm20(stat)\pm87(syst)\,MeV$
      without HT.
\item The illustrative nature of the QCD fit to the data on $F_2$
      should be pointed out. The matter is the absence
      of reliable theoretical predictions for HT contribution to
      singlet SF. In spite of this, we obtained the momentum fraction
      carried by quarks in the nucleon, ~$M_q^{QCD}(2,Q^2)=0.46$,
      to be in agreement with the previous measurements.
\item The strong interaction constant at the point of Z boson mass is
      found to be higher than most of the deep inelastic scattering
      results~\cite{Shif,Bethke}.
\item The consideration of the HT contribution decreases
      $\chi ^2$ and  appreciably  changes the parameters of the fit
      as well as the GLS sum rule value and $\alpha_S (M_Z)$. For a reliable
      QCD analysis one must calculate not only the GLS sum rule
      ($N=1$) but also the higher SF moments ($N=2,3,...$).
      Using in addition a 3-loop QCD analysis one could expect
      to improve the estimation of $\alpha_S (M_Z)$.
\end{itemize}


In conclusion, let us stress once more that the QCD analysis
of SF is sensitive to the HT contribution and in the future it should
take into account the nuclear effects, heavy quark
threshold effects and higher order QCD corrections.
We hope to improve the accuracy of our estimations by processing the
additional data on deep inelastic scattering obtained with the IHEP--JINR
Neutrino Detector in the wide band beams of $\nu_\mu$ and $\overline\nu_\mu$.

\newpage
\begin{center}
{\small{
\begin{tabular} {lccc}
\hline
Beam  & ~~$\overline\nu_\mu$~ &
                ~$\nu_\mu$~ & ~~$\nu_\mu$~ \\
$E_p\,(GeV)$~~ & ~~$70$~ & ~$70$~ & ~$67$~\\
Final statistics~~ & ~~$741$~ & ~$2139$~ & ~$3848$~\\
$\langle Q^2 \rangle\,(GeV^2)$~~ & ~$1.2$ &
                 \multicolumn{2}{c}{~$2.3$}\\
\hline
\end{tabular}
}}
\end{center}
Table~1.~~{Summary of the exposures.}

\newpage
\begin{center}
{\small{
\begin{tabular}{c|c|rccc|ccc}
\hline
$x$ & $\langle Q^2\rangle\,(GeV^2)$ & $F_2$~~ & $stat$ & $syst$ &
            $\Delta F_2$~ & $xF_3$ & $stat$ & $syst$ \\
\hline
~$.052$~ & $.55$ & ~$1.169$ & ~$.026$~ & ~$.047$~ & ~$.023$~~ &
                   ~$.445$~ & ~$.044$~ & ~$.062$~ \\
~$.148$~ & $1.4$ &  $1.097$ & $.026$ & $.022$ & $.022$  &
                    $.583$  & $.044$ & $.017$ \\
~$.248$~ & $2.2$ &  $.894$  & $.023$ & $.018$ & $.019$  &
                    $.622$  & $.038$ & $.019$ \\
~$.346$~ & $2.9$ &  $.576$  & $.016$ & $.017$ & $.013$  &
                    $.556$  & $.027$ & $.011$ \\
~$.447$~ & $3.4$ &  $.390$  & $.014$ & $.012$ & $.009$  &
                    $.336$  & $.023$ & $.007$ \\
~$.563$~ & $4.0$ &  $.182$  & $.008$ & $.004$ & $.004$  &
                    $.177$  & $.012$ & $.005$ \\
\hline
\end{tabular}
}}
\end{center}
Table 2.~~
The isoscalar structure functions ~$F_2$~ and
          ~$xF_3$~ obtained on the assumption of ~$R=0$. The difference
          ~$\Delta F_2$~ between the values of ~$F_2$~ obtained
          with ~$R=.1$~ and those obtained with ~$R=0$~ is
          also presented. The bin edges are at
          ~$x=.0, ~.1, ~.2, ~.3, ~4, ~.5, ~.65$.
\ \\
\ \\
\ \\
\ \\
\newpage
\begin{center}
{\small{
\begin{tabular} {ccc}        \hline
                              &   $
\langle\langle O\rangle\rangle=0$             &  $
\langle\langle O\rangle\rangle=0.53$ \cite{HT}  \\      \hline
     $\chi^2$                 & 2.8                   &2.05                 \\
        A                     & 9.28    $ \pm $  1.73 &0.90  $ \pm $   0.67 \\
        b                     & 1.06    $ \pm $ 0.11  &0.31  $ \pm $   0.18 \\
        c                     & 3.22    $ \pm $ 0.31  &3.64  $ \pm $   0.21 \\
    $\gamma$                  &-0.90    $ \pm $ 0.21  &9.53  $ \pm $  5.73  \\
$\Lambda_{\overline{MS}}$~[MeV] &   417 $ \pm $   51  &358   $ \pm $  59    \\
GLS sum rule                  & 1.59                         & 2.63  \\
$\alpha_S (M_Z)$              &$0.123^{+3}_{-4} $ & $0.120^{+3}_{-4}$   \\
\hline
\end{tabular}
}}
\end{center}
Table 3.~~The results of the NLO QCD fit
to the $xF_3$ SF data for $f=4$, $Q^2_0\,=\,3\,GeV^2$, $N_{MAX}=12$,
$\alpha~=0.7$, $\beta~=3.0$ with the corresponding statistical errors.
\newpage
Figure captions. \\[5mm]

Fig.1.
The $x$-dependence of the isoscalar
         structure functions $F_2(x)$ and $xF_3(x)$.
          The statistical and systematic
          errors are added in quadrature, excluding the
          normalization error of ~$4\%$~ for ~$F_2$~ and
          ~$11\%$~ for ~$xF_3$. The curve is fit
          of the form ~$xF_3(x)=A\,x^{\,b}(1-x)^{\,c}$.
          ~The best fit values of free parameters
          ~$A=5.36\pm 1.25\,(stat)$, ~$b=0.81\pm 0.10\,(stat)$,
          ~$c=3.52\pm 0.26\,(stat)$~ were obtained using for
          each $x$-bin the mean ~$x$~ of the bin as the
          actual $x$-point corresponding to the value of the
          structure function obtained.

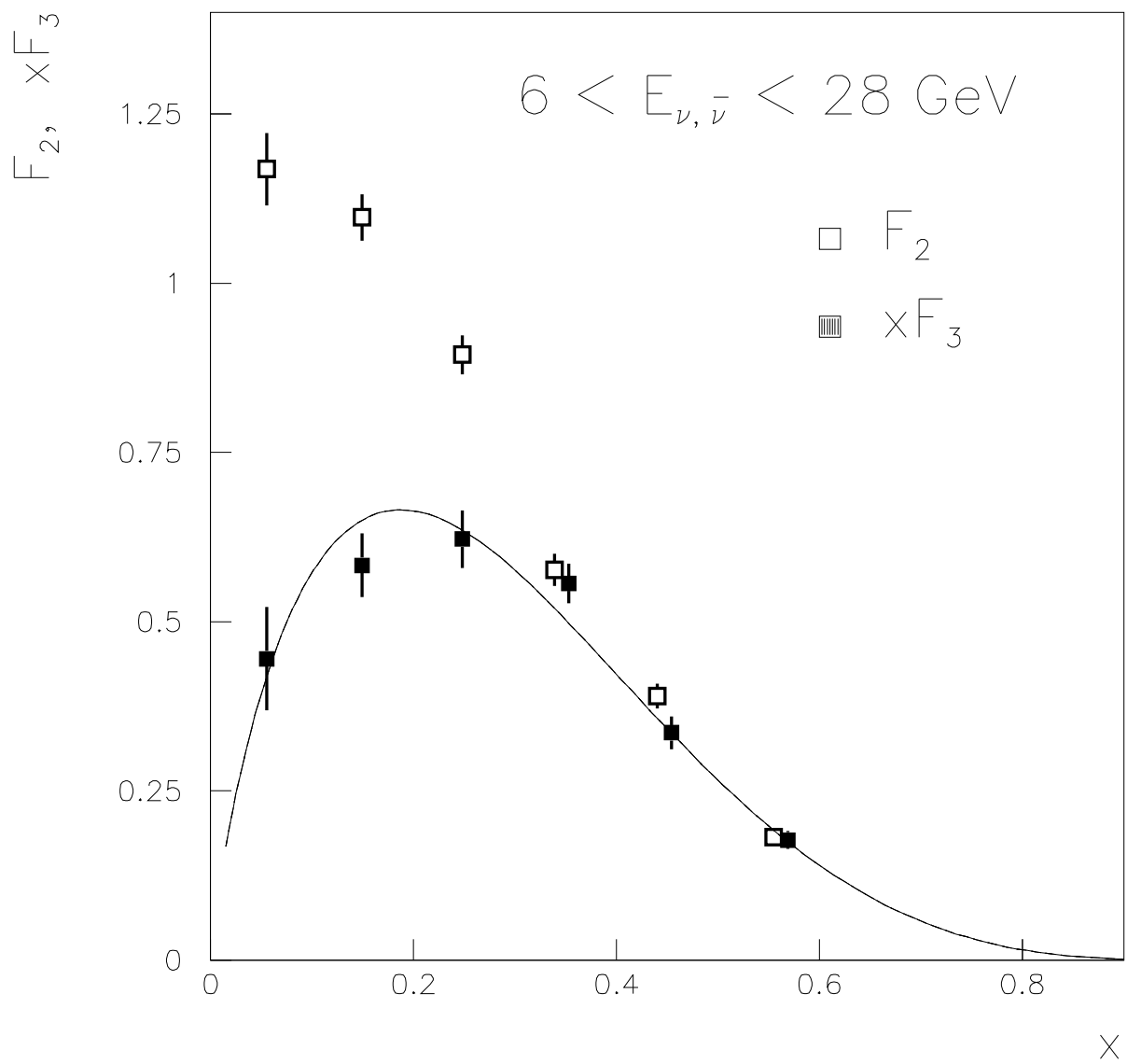
\end{document}